\documentclass[prl,showpacs,letterpaper,preprint,groupedaddress]{revtex4}

\usepackage{helvet}
\usepackage{amsmath}
\usepackage{amssymb}
\usepackage{bm}
\usepackage{bbm}
\usepackage{units}
\usepackage{hyperref}
\usepackage{graphicx}

\newcommand{\tensorhelv}[1]{\fontfamily{phv}\selectfont\textbf{#1}}

\begin{document}
\title{Parametric Amplification in Dynamic Radiation Force of Acoustic Waves}
\author{Glauber T. Silva$^1$} 
\email{glauber@tci.ufal.br}
\author{Shigao Chen$^2$}
\author{Leonardo P. Viana$^1$}
\affiliation{$^1$Instituto de Computa\c{c}\~ao, Universidade Federal de Alagoas, Macei\'o, AL, Brasil, 57072-970}
\affiliation{$^2$Department of Physiology and Biomedical Engineering, Mayo Clinic College of Medicine, Rochester, MN, USA, 55905}

\begin{abstract}
We report on parametric amplification in dynamic radiation force produced by a bichromatic acoustic beam in a fluid.
To explain this effect we develop a theory taking into account the nonlinearity of the fluid.
The theory is validated through an experiment to measure the dynamic radiation force on an acrylic sphere.
Results exhibit an amplification of $\unit[66]{dB}$ in water and $\unit[80]{dB}$ in alcohol as the difference of the frequencies is increased from $\unit[10]{Hz}$ to $\unit[240]{kHz}$.
\end{abstract}

\maketitle

Acoustic radiation force in fluids is a phenomenon that has been investigated for over a century~\cite{rayleigh02a}.
It results from variations in energy and momentum of the wave as a consequence of scattering, attenuation, or distortion by nonlinear effects.
This force is similar to the optical radiation pressure exerted by electromagnetic waves on responsive particles~\cite{ashkin70a}.
Radiation force can be either static or dynamic with respect to its time dependency.
Static radiation force is a time-averaged quantity produced by a monochromatic wave that corresponds to the dc component in the spectrum of the stress (pressure)~\cite{borgnis53b}.
Dynamic radiation force can be generated by a bichromatic acoustic wave.
This force is associated to the difference frequency component in the spectrum of the stress~\cite{silva2005a}.

One of the first applications of dynamic radiation force was devised in $1928$ by Sivian~\cite{sivian28a}  to measure acoustic power on a suspended disk.
In $1953$, Macnamara \textit{et al.}~\cite{mcnamara52a} developed a method based on this force to measure absorption in liquids.
After these pioneer applications, dynamic radiation force has passed unnoticed until late 1970's. 
In the last three decades, it has been applied for measuring ultrasound power of transducers~\cite{greenspan78a}, inducing oscillation in bubbles~\cite{marston80a} and liquid drops~\cite{marston:1233}, and exciting modes in capillary bridges~\cite{morse1996}.
Furthermore, dynamic radiation force is the underlying principle of some elastography imaging techniques such as shear wave elasticity imaging~\cite{sarvazyan98a} and vibro-acoustography~\cite{fatemi98a}. 

Despite early applications, dynamic radiation force was only recently investigated in theoretical grounds. 
Mitri~\textit{et al.}~\cite{mitri2005} calculated it on elastic cylinders.
He also studied the force produced by a bichromatic standing plane wave~\cite{mitri2005D}.
Silva~\textit{et al.}~\cite{silva2005a} obtained the dynamic radiation force on elastic spheres which was confirmed in an experiment realized by Chen~\textit{et al.}~\cite{chen05a}.
We emphasize these results are only valid when the difference between the fundamental frequencies of the wave is very narrow.
So far, no one has taken into account the influence of the fluid nonlinearity in dynamic radiation force thoroughly.
The nonlinearity of the fluid is described by the thermodynamic relation $p\propto \rho^{(1+B/A)}$, where $p$ and $\rho$ are respectively the pressure and the density of the fluid, and $B/A>0$ is commonly used in acoustics as the nonlinear parameter~\cite{fox54a}. 
It is worthy to note that static radiation force in a ideal fluid does not depend on the nonlinearity parameter $B/A$~\cite{beissner95a}. 
On the other hand, one may inquire on how does dynamic radiation force depend on the nonlinearity of the fluid?

In this letter we undertake this question.
Our analysis unfolds that dynamic radiation force may achieve a regime of parametric amplification.
The concept of parametric amplification arose in radio engineering and is widely known in optics~\cite{yariv1989}.
In acoustics, parametric amplification can also be understood as follows.
The mixture of two waves of differing angular frequencies $\omega_1$ and $\omega_2$ $(\omega_2>\omega_1)$ generates two new waves; one of which has frequency equal to $\omega_1 + \omega_2$, while the other arises with the difference frequency $\omega_{21}=\omega_2 - \omega_1$~\cite{thuras1935,westervelt63}.
To demonstrate the parametric regime in dynamic radiation force,
we calculate this force on a rigid sphere taking into account the nonlinearity of the fluid.
An experiment using a laser vibrometer is designed to measure the dynamic radiation force on an acrylic sphere immersed in degassed water and ethyl alcohol.
Results show an amplification of this force of up to $\unit[66]{dB}$ in water and $\unit[80]{dB}$ in alcohol as the difference frequency varies from $\unit[10]{Hz}$ to $\unit[240]{kHz}$.

Consider a homogeneous and isotropic fluid with adiabatic speed of sound $c_0$, in which thermal conductivity and viscosity are neglected. 
The fluid has infinite extent and is characterized by the following acoustic fields: pressure $p$, density $\rho$, and particle velocity $\mathbf{v}=-\nabla \phi$.
The function $\phi$ is the velocity potential and $\nabla$ is the gradient operator. 
These fields are function of the position vector $\mathbf{r}$ and time $t$.
At rest, these quantities assume constant values
$p=p_{0}$, $\rho = \rho_{0}$, and $\mathbf{v}=0$.
The acoustic fields are governed by the dynamic equations of ideal fluids.
By using the regular perturbation technique, one can expand the velocity potential in terms of the Mach number $\varepsilon\ll 1$ as
$\phi = \varepsilon \phi^{(1)} + \varepsilon^2 \phi^{(2)}+O(\varepsilon^3).$
Any analysis of radiation force has to be done considering at least the second-order terms of this expansion.

The excess pressure in the fluid can be written as $p-p_0 = p^{(1)}+p^{(2)}+O(\varepsilon^3)$, where~\cite{lee93a} 
\begin{equation}
\label{acoustic_pressure}
p^{(1)} =  \varepsilon \rho_0 \frac{\partial \phi^{(1)}}{\partial t}
\end{equation}
and
\begin{equation}
\label{2nd_pressure}
p^{(2)} = \varepsilon^2 \rho_0 \left[\frac{1}{2c_0^2} \left(\frac{\partial \phi^{(1)}}{\partial t}\right)^2 -\frac{\|\nabla \phi^{(1)}\|^2}{2}  +  \frac{\partial \phi^{(2)}}{\partial t}\right]
\end{equation}
are the acoustic and the second-order pressure fields, respectively.
The first two terms in the right-hand side of Eq.~(\ref{2nd_pressure}) depend only on $\phi^{(1)}$.
They correspond to the Lagrangian density of the wave.
As we shall see the potential $\phi^{(1)}$ does not depend on the nonlinearity of the fluid, while $\phi^{(2)}$ does.
Thus, Eq.~(\ref{2nd_pressure}) has contribution of two terms called here the Lagrangian and the nonlinear pressures.

Let $S_0$ be the surface of the object target at rest.
One can show that the instantaneous force on the object up to second-order in the excess of pressure is given by
\begin{equation}
\label{rad_force1}
\mathbf{f} = -\iint_{S(t)} p^{(1)}\mathbf{n}dS -\iint_{S_0} p^{(2)} \mathbf{n} dS, 
\end{equation}
where $S(t)$ is the moving object surface, $\mathbf{n}$ is the outward normal unit-vector on the integration surface.
Assuming that the sphere is under influence of a bichromatic acoustic beam with fundamental angular frequencies $\omega_1$ and $\omega_2$, the dynamic radiation force is produced by the contribution of stresses at the difference frequency $\omega_{21}=\omega_2-\omega_1$.

Consider the Fourier transform of a function $g(t)$ as $\mathcal{F}[g]$ and its inverse denoted by $\mathcal{F}^{-1}$.
The dynamic radiation force is given in terms of the component of Eq.~(\ref{rad_force1}) at $\omega_{21}$. 
Accordingly, the dynamic radiation force is~\cite{silva2005a}
$\mathbf{f}_{21} = \mathcal{F}^{-1}\left[\hat{\mathbf{f}}_{21}\right]$, 
where
\begin{equation}
\label{f21}
\hat{\mathbf{f}}_{21} = \iint_{S_0}\mathbf{n} \cdot \hat{\tensorhelv{S}}_{21} dS  - i\varepsilon\rho_0\omega_{21}\mathcal{F}\iint_{S(t)} \phi^{(1)} \mathbf{n}dS\biggr|_{\omega_{21}},
\end{equation}
where $i$ is the imaginary unit and 
\begin{equation}
\hat{\tensorhelv{S}}_{21} = -\mathcal{F}[p^{(2)}\tensorhelv{I} + \rho_0 \mathbf{v}^{(1)}\mathbf{v}^{(1)}]_{\omega_{21}}
\end{equation} 
is the amplitude of the dynamic radiation stress with $\tensorhelv{I}$ being the unit tensor.
The dyad $\rho_0 \mathbf{v}^{(1)}\mathbf{v}^{(1)}$ is the Reynolds' stress tensor.

To obtain the dynamic radiation force over an object, we have to solve the corresponding scattering problem described by $\phi^{(1)}$ and $\phi^{(2)}$.
These functions satisfy the linear and the second-order wave equations~\cite{heaps:355}
\begin{eqnarray}
\label{phi1}
\square^2 \phi^{(1)} &=& 0, \\
\square^2 \phi^{(2)} &=&
-\frac{1}{c_0^2}\frac{\partial}{\partial t}\biggl[ \frac{1}{2}\square^2 \phi^{(1)2}  
+ \frac{\gamma}{c_0^2}\left(\frac{\partial \phi^{(1)}}{\partial t}\right)^2\biggr], 
\label{phi2}
\end{eqnarray}
where $\square^2 = \nabla^2 - (1/c_0^2)(\partial/\partial t)^2$ is the d'Alembertian operator and $\gamma = 1 + B/A$.
Hence, the scattering problem should be solved through Eqs.~(\ref{phi1}) and (\ref{phi2}) with appropriate boundary conditions.

Now consider a bichromatic plane wave formed by the excitation in the velocity field $\mathbf{v} = \varepsilon c_0 (\sin \omega_1 t + \sin \omega_2 t)\mathbf{e}_z$ at $z=0$, where $\mathbf{e}_z$ is the unit vector in the  $z$ direction.
For the perturbation terms of the velocity potential, one has the boundary conditions $\partial \phi^{(1)}/\partial z=-c_0(\sin\omega_1 t + \sin\omega_2 t)$ and $\partial \phi^{(2)}/\partial z = 0$ both at $z=0$. 
The solution of Eq.~(\ref{phi1}) is  
\begin{equation}
\label{phi1_s}
\phi^{(1)} = c_0\text{Re}\left\{\frac{1}{k_1}e^{-i(\omega_1 t - k_1 z)} + \frac{1}{k_2}e^{-i(\omega_2 t - k_2 z)}\right\},
\end{equation}
where $\text{Re}$ means the real-part of a complex variable, $k_1=\omega_1/c_0$, and $k_2=\omega_2/c_0$.
We are only interested in the second-order velocity potential at the difference frequency $\omega_{21}$.
Hence, from Eq.~(\ref{phi2}) we have 
\begin{equation}
\label{phi_21} 
\phi^{(2)}_{21} = \frac{\gamma c_0}{2 k_{21}} \text{Re}\left\{\left( k_{21}z + i  \right) e^{-i(\omega_{21}t - k_{21}z)}\right\},
\end{equation}
where $k_{21}=\omega_{21}/c_0$.

In what follows we calculate the dynamic radiation force on a rigid sphere of radius $a$.
The amplitude of the force in Eq.~(\ref{f21}) has contributions from the Lagrangian, the acoustic and the nonlinear pressures, and the Reynolds' stress tensor. 
One can show that the contribution of the second term in the right-hand side of Eq.~(\ref{f21}) is proportional to $(\omega_m Z_m)^{-1}$, $m=1,2$.
The quantity $Z_m$ is the mechanical impedance of the oscillating sphere~\cite{chen05a} at the frequencies $\omega_1$ and $\omega_2$.
In the experimental setup we use frequencies above $\unit[2.2]{MHz}$; hence, the magnitude of $(\omega_m Z_m)^{-1}$ is as small as $10^{-6}$, which is much smaller than other contributions to the dynamic radiation force.
Therefore, we neglect this contribution here.

Let us focus on the scattering of the nonlinear pressure at the difference frequency $p_{21}=\varepsilon^2 \rho_0 \partial \phi^{(2)}_{21}/\partial t$.
We restrict our analysis to the case $k_{21} z_0 \gg 1$ and $a \ll z_0$.
The energy density at the acoustic source is $E_0= \varepsilon^2 \rho_0 c_0^2/2$.
From Eq.~(\ref{phi_21}), the amplitude of the nonlinear pressure in the spherical coordinates $(r,\theta,\varphi)$, is given in terms of partial spherical waves as 
\begin{equation}
\hat{p}_{21} = A\sum_{n=0}^{+\infty}(2n+1) i^n [j_n(k_{21}r) + b_n h_n(k_{21}r)] P_n(\cos \theta), 
\label{phi_sphere}
\end{equation}
where 
$A = -i E_0 \gamma k_{21}z_0 e^{ik_{21}z_0}$, the functions $j_n$ and $h_n$ are the spherical Bessel and first-type Hankel functions of $n$th-order, respectively.
The function $P_n$ is the Legendre polynomial of $n$th-order.
The scattering coefficients are given by~\cite{faran51a} 
\begin{eqnarray*}
b_n &=& - j'_n(x_{21}) / h'_n(x_{21}), \quad n\neq1  \\
b_1 &=& - \frac{(\rho_0/\rho_1)j_1(x_{21}) - x_{21}j'_1(x_{21})}
{(\rho_0/\rho_1)h_1(x_{21}) - x_{21} h_1'(x_{21})},
\end{eqnarray*}
where $x_{21} = k_{21}a$.
The symbol $'$ stands for the derivative.

By substituting Eq.~(\ref{phi_sphere}) into Eq.~(\ref{f21}), neglecting the direct contribution of the acoustic pressure, and using the result of Ref.~\cite{silva2005a}, one obtains the dynamic radiation force as
\begin{equation}
\label{f21_2}
\mathbf{f}_{21} = \pi a^2 E_0 \hat{Y}_{21} e^{-i (\omega_{21}t- k_{21}z_0)} \mathbf{e}_z.
\end{equation}
The dynamic radiation force function $\hat{Y}_{21}$ is given by
\begin{eqnarray}
\nonumber
\hat{Y}_{21} &=&  -\frac{4}{x_1 x_2}\biggl\{ 
\nonumber
\frac{6\rho_0}{\rho_1} \left( R^*_{1,1} R_{2,2} + R_{1,2}^*R_{2,1}\right)
+\sum_{n=0}^{+\infty} (n+1)\\
\nonumber
&\times& \left[x_1 x_2 - n(n+2) \right]
 \left( R_{1,n}^* R_{2,n+1} + R^*_{1,n+1} R_{2,n} \right)
\biggr\}  \\
&+& \frac{i(z_0/a)\gamma}{(\rho_0/\rho_1)h_1(x_{21})-x_{21} h'_1(x_{21})},
\label{Y}
\end{eqnarray}
where
\begin{eqnarray}
\nonumber
R_{m,n} &=& \frac{i^{n+1}}{x_m^2 h'_n(x_m)}, \quad n\neq 1,\\
\nonumber
R_{m,1} &=& \frac{1}{x_m\left[ (\rho_0/\rho_1)h_1(x_m) - x_m h_1'(x_m)\right]},
\end{eqnarray}
$x_m=k_m a$ with $m=1,2$.
The symbol $\mbox{}^*$ is the complex conjugate.
The first term of Eq.~(\ref{Y}) is the dynamic radiation force used in the literature~\cite{mitri2005,silva2005a}.
The last term in the right-hand side of Eq.~(\ref{Y}) is due to the nonlinearity of the fluid and is associated to parametric amplification in dynamic radiation force. 
This term has not been treated in previous works.
The regime of parametric amplification can only be neglected  when $\omega_1 \simeq \omega_2$.
It is important to note this amplification depends on the quantities, namely, the ratio $z_0/a$, $\gamma$, and $\omega_{21}$.
\begin{figure}[t]
\begin{center}
	\includegraphics[width=\linewidth]{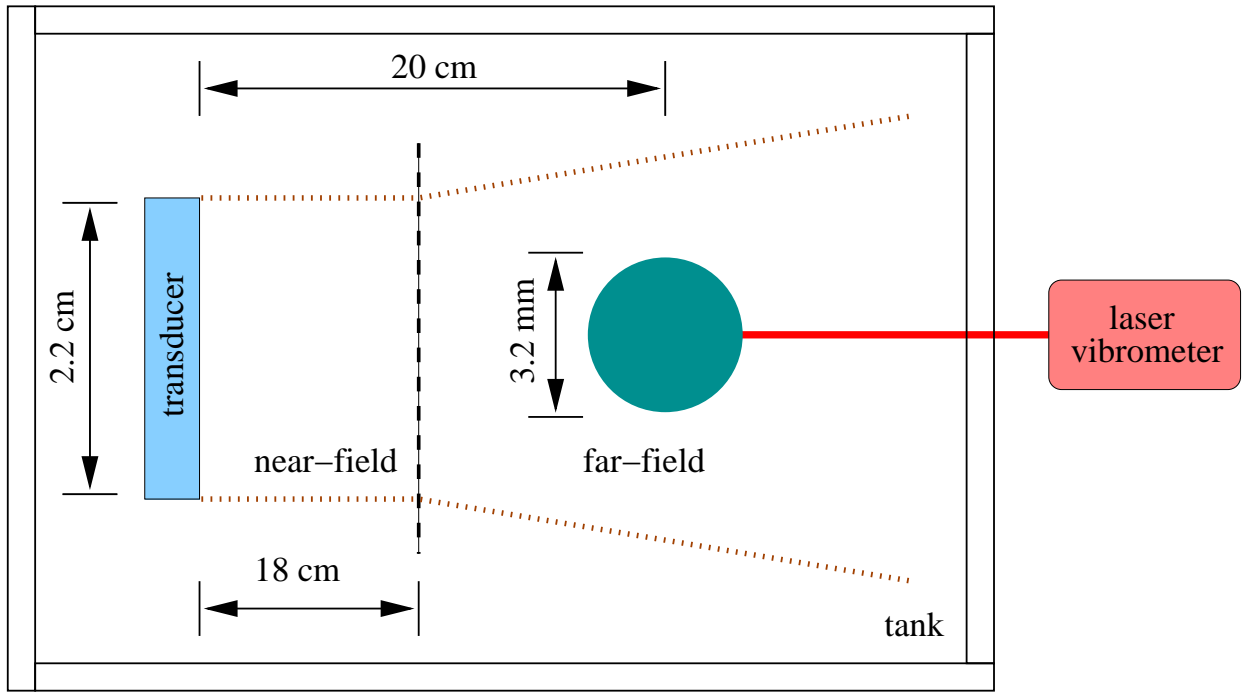}
\end{center}
\caption{(color online) Experimental apparatus utilized to measure the dynamic radiation force on the sphere.
The dimensions of the tank is $100\times64\times\unit[38]{cm}$ for water and $40\times40\times\unit[24]{cm}$ for alcohol.
}
\label{fig:experiment}
\end{figure}

To verify Eq.~(\ref{f21_2}), we realized an experiment to measure dynamic radiation force.
The basic experimental setup is described in detail in Ref.~\cite{chen05a}.
In Fig.~\ref{fig:experiment} we illustrate the experimental apparatus.
An in-house flat transducer with diameter $d=\unit[22]{mm}$ is used to insonify an acrylic sphere inside a tank filled either ethyl alcohol or degassed water.
The sphere has radius $a=\unit[1.6]{mm}$, density $\rho_1 = \unit[1190]{Kg/m^3}$ and
is suspended in a pendulum along the beam axis $\unit[20]{cm}$ away from the transducer.
The parameters for alcohol (water) are $\rho_0 = 785~\unit[(1000)]{kg/m^3}$, $c_0 = 1100~\unit[(1500)]{m/s}$, and $\gamma=11~(6)$.
We used a suppressed-carrier amplitude-modulated signal to drive the transducer.
The carrier frequency is $\omega_0/2\pi=\unit[2.25]{MHz}$ and the modulation frequency $\omega_{21}/2\pi$ is swept from $\unit[10]{Hz}$ to $\unit[240]{kHz}$. 
In such configuration the wave frequencies are $\omega_{1,2} = \omega_0 \mp \omega_{21}/2$.
For the specified difference frequency range, we measured the acoustic pressure in water at $\unit[200]{mm}$ away from the transducer aligned with the beam axis.
The measurement was performed by a polyvinylidene fluoride membrane hydrophone (Y-33-7611, GEC-Marconi, Great Britain).
The amplitude of the measured pressure remained constant within an error of less than $5\%$ as the difference frequency varied in the specified range.
The dynamic radiation force on the sphere is obtained by measuring vibration velocity of the sphere.
The force is given by the product of the vibration velocity and the mechanical impedance of the sphere at $\omega_{21}$.
The vibration of the sphere was detected by a laser vibrometer (Polytec GmbH, Waldbronn, Germany), which was aligned with the beam axis of the transducer.
The signal from the vibrometer was filtered by a lock-in amplifier (Perkin Elmer 7265, Oak Ridge, TN) at the difference frequency $\omega_{21}$.

The frequencies used in the experiment are such that $k_{1,2} a \gg 1$.
Thus the distance at which the farfield of the transducer begins for the frequencies $\omega_{1}$ and $\omega_{2}$ is~\cite{kino87a} $z_{1,2} = k_{1,2}d^2/8\pi$, which corresponds to $z_1 \simeq z_2 = \unit[18]{cm}$.
Furthermore, in this region the incident wave resembles a plane wave with circular cross-section whose $\unit[3]{dB}$-radius is approximately~\cite{kino87a} $0.35 d/2 = \unit[3.9]{mm}$.
We can reasonably assure that the sphere is thoroughly inside this region and the plane wave approximation can be used for the incident beam.

The dynamic radiation force as described in Eq.~(\ref{f21_2}) is only valid within the preshock wave range.
When a finite-amplitude wave propagates in a fluid, its form tends to steepen up to develop shocks.
The formation of shock waves implies dissipation, which is not described by Eqs.~(\ref{phi1}) and (\ref{phi2}).
For a monochromatic plane wave the preshock range has length~\cite{hamilton98a}
$\ell = \lambda[2\pi \varepsilon (1+B/2A)]^{-1},$ where $\lambda$ is the wavelength.
The only unknown parameter necessary to determine the preshock wave range is the Mach number.
However, we may estimate it by equating the dynamic radiation force measured at $\omega_{21}/2\pi = \unit[10]{Hz}$ to Eq.~(\ref{f21_2}).
In this frequency range the theory of dynamic radiation force was already validated experimentally~\cite{chen05a}.
A summary of the measured data in the radiation force experiment is shown in Table~\ref{table:exp_summary}.
According to the measured data, the preshock wave range is $\ell = 62~\unit[(272)]{cm}$ for alcohol (water).
Therefore, the sphere at $z_0=\unit[20]{cm}$ is inside this range for both alcohol and water experiments.
\begin{table}[t]
\caption{Summary of the experimental measurements.}
\label{table:exp_summary}
\begin{tabular}{lcccc}
\hline
& Mach &\multicolumn{2}{c}{Dynamic Radiation Force} & Gain\\
\cline{3-4}& Number & $\unit[10]{Hz}$ & $\unit[240]{kHz}$   & \\ 
\hline
\hline 
alcohol & $1.7\times10^{-5}$  & $\unit[3.0]{\mu N}$& $~\unit[34]{m N}$&  $\unit[80]{dB}$ \\ 
\hline 
water   & $1.0\times10^{-5}$ & $\unit[2.8]{\mu N}$& $\unit[6.7]{m N}$& $\unit[66]{dB}$ \\ 
\hline 
\end{tabular} 
\end{table}

\begin{figure}[t]
\begin{center}
	\includegraphics[width=\linewidth]{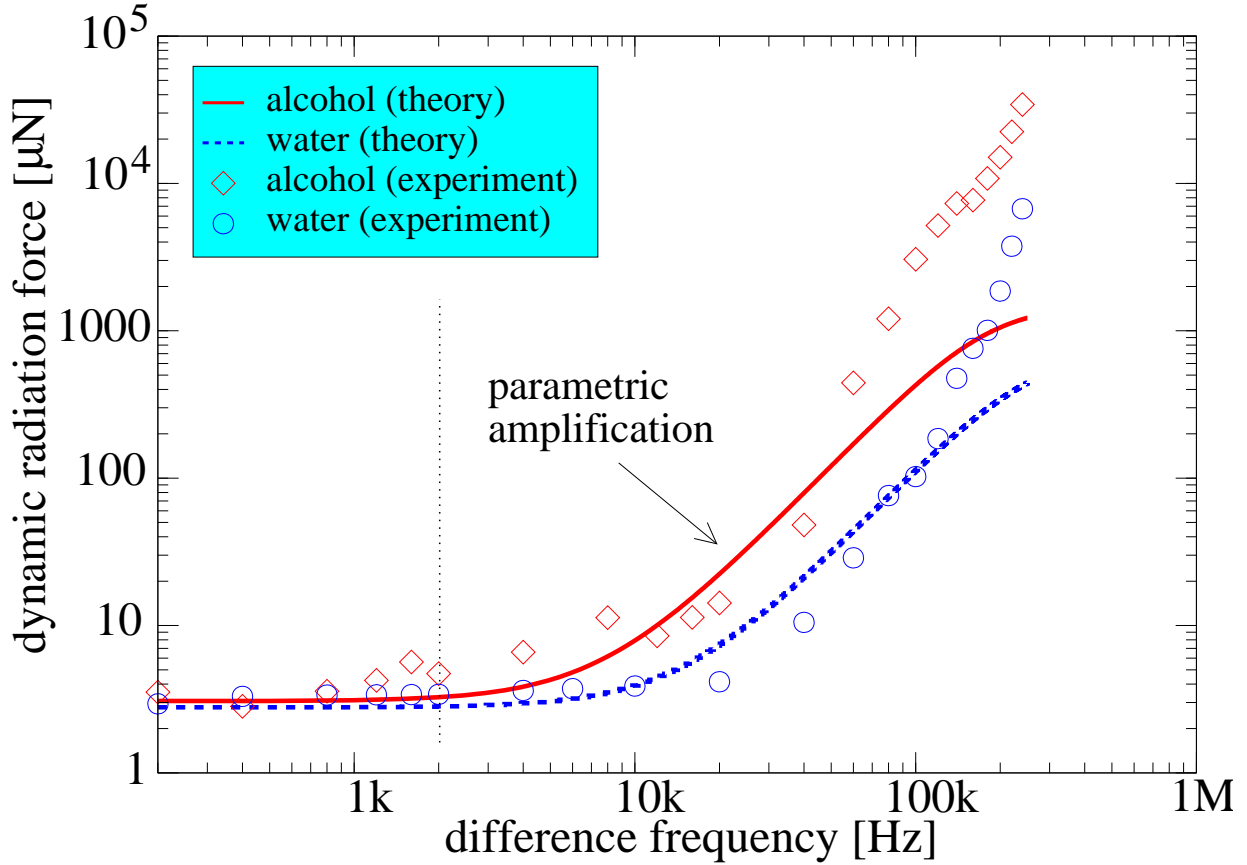}
\end{center}
\caption{(color online) Comparison of the theory and experiment to measured dynamic radiation force on the acrylic sphere suspended in water $(\circ)$ and ethyl alcohol $(\diamond)$.} 
\label{fig:sphere}
\end{figure}

In Fig.~\ref{fig:sphere}, we exhibit a comparison between the measured dynamic radiation force and Eq.~(\ref{f21_2}).
Prior to the vertical dotted line at $\unit[2]{kHz}$, parametric amplification has a minor role in the dynamic radiation force. 
This region coincides with the result obtained in Ref.~\cite{silva2005a}. 
Beyond this line we have a prominent amplification of the dynamic radiation force.
The theory predicts a gain of $52~\unit[(44)]{dB}$ in alcohol (water) as the difference frequency varies from $\unit[10]{Hz}$ to $\unit[240]{kHz}$.
Discrepancies between the theory and experimental results might be related to diffraction, thermoviscous effects of the fluid, and surface and elastic properties of the sphere.
Factually, the acrylic sphere is a viscoelastic material which allows the formation of internal and surface waves.
These waves are not present in a rigid sphere.
Thermoviscous effects may produce streaming in boundary layers surrounding the sphere and nearby the walls of the tank.
Streaming can cause a blueshift in the waves near to the surface of the sphere; thus increasing the dynamic radiation force.
A theory with thermoviscous effects requires a full solution of the scattering problem using the Navier-Stokes equations~\cite{doinikov96a}.

In summary, we showed that dynamic radiation force is subjected to parametric amplification.
Measured amplification were $\unit[80]{dB}$ in alcohol and $\unit[66]{dB}$ in water as the difference frequency was tuned from $\unit[10]{Hz}$ to $\unit[240]{kHz}$.
Results are in reasonable agreement with the theory presented in this letter.
We believe parametric amplification may set new applications of radiation force in elastography and trapping particles.

The authors thank F. Mitri, J. F. Greenleaf, and M. Fatemi for helpful discussions.
Work partially supported by FAPEAL and CNPq (Brazilian agencies).

\bibliographystyle{apsrev}
\bibliography{mybib}

\end{document}